# First-principles Study on the Magnetic Interactions in Honeycomb Na$_2$IrO$_3$

Y. S. Hou[1,2], J. H. Yang[3], H. J. Xiang[1,2,*], and X. G. Gong[1,2,*]

[1] Key Laboratory of Computational Physical Sciences (Ministry of Education), State Key Laboratory of Surface Physics, Department of Physics, Fudan University, Shanghai 200433, China

[2] Collaborative Innovation Center of Advanced Microstructures, Nanjing, 210093, China

[3] National Renewable Energy Laboratory, Golden, Colorado 80401, United States

Email: hxiang@fudan.edu.cn, xggong@fudan.edu.cn

**Abstract**

Honeycomb iridate Na$_2$IrO$_3$, a $J_{eff}$=1/2 magnet, is a potential platform for realizing the quantum spin liquid. Many experiments have shown that its magnetic ground state has a zigzag antiferromagnetic (AFM) order. However, there is still a lack of consensus on the theoretical model explaining such order, since its second nearest neighbor (NN) and long-range third NN magnetic interactions are highly unclear. By properly taking into account the orbital moments, achieved through constraining their directions in the first-principles calculations, we obtain that the relative angle between orbital and spin moments is fairly small and in the order of several degrees, which thus validates the $J_{eff}$=1/2 state in Na$_2$IrO$_3$. Surprisingly, we find that the long-range third NN Heisenberg interactions are sizable whereas the second NN magnetic interactions are negligible. Using maximally localized Wannier functions, we show that the sizable long-range third NN Heisenberg interaction results from the extended nature of the $J_{eff}$=1/2 state. Based on our study, we propose a minimal $J_1$-$K_1$-$\Gamma_1$-$J_3$ model in which the magnetic excitations have an intensity peak at 5.6 meV, consistent with the inelastic neutron scattering experiment [Phys. Rev. Lett. **108**, 127204 (2012)]. The present work demonstrates again that constraining orbital moments in the first-principles calculations is powerful to investigate the intriguing magnetism in the $J_{eff}$=1/2 magnets, and paves the way toward gaining a deep insight into the novel magnetism discovered in the honeycomb $J_{eff}$=1/2 magnets.

**PACS number(s):** 75.10.Jm, 75.50.Ee, 02.70.Uu

## I. INTRODUCTION

Recently, iridium oxides had been increasingly studied both experimentally and theoretically. Many novel and intriguing phenomena have been put forward [1], such as topological Mott insulator [2], Weyl semimetal and axion insulator [3], unconventional high-temperature superconductivity [4, 5], *etc*. Among the iridium oxides, the honeycomb $A_2$IrO$_3$ ($A$=Li, Na) are of particular interest because it has been theoretically predicted [6, 7] that they could realize the long-sought Kitaev model, which has an exactly solvable quantum spin liquid ground state [8]. Many experimental studies on these iridium oxides are also inspired to discover the exotic and interesting quantum spin liquid [9-14].

Experimentally, it has been shown that the prototypical honeycomb iridium oxide Na$_2$IrO$_3$ has a zigzag AFM order and its magnetic easy axis is approximately half way between the cubic *x*- and *y*-axes [9, 10, 14, 15]. Theoretically, a general consensus among the models explaining such order is still lacking and there exist many diverse models [7, 11, 12, 16-24]. The most controversial issue focuses on the second and third NN magnetic interactions. Firstly, models consisting of only the NN interactions are proposed to give rise to the zigzag AFM order [7, 17, 21], whereas it is urged that the second and third NN magnetic interactions should also be taken into account [11, 12, 18-20]. Secondly, the types and the strengths of the second and third NN magnetic interactions are strikingly disputed. Kimchi *et al*. proposed that the zigzag AFM order could be explained by the Heisenberg-Kitaev model only plus the second and third NN Heisenberg interactions (HK-$J_1$-$J_2$ model) [18], but the second NN Kitaev interaction is also argued to be important [19]. As for their strength, fitting the HK-$J_1$-$J_2$ model to the experimentally measured spin wave shows $J_2/J_1$=0.78 and $J_3/J_1$=0.9 [11], but theoretical calculations using nonperturbative exact diagonalization methods demonstrate that the long-range third NN Heisenberg interaction $J_3$ is unexpectedly strong while both the NN Heisenberg interaction $J_1$ and the second NN Heisenberg interaction $J_2$ are negligible [20]. These results are elusive because the bond distances of both the second NN and the third NN Ir-Ir pairs are nearly two times longer than those of the NN Ir-Ir pairs.

To gain a deeper insight into the zigzag AFM order in Na$_2$IrO$_3$, the key is to determine the magnetic interactions fully, especially the disputed second NN and the long-range third NN on the same footing. So far there are only several first-principles calculations which estimated the second NN and the third NN magnetic interactions [23, 24], although the NN magnetic interactions have been thoroughly investigated by different methods in many previous studies [19, 22-25]. An essential difficulty lies in that orbital moments play an important role in determining the magnetic interactions in the $J_{\text{eff}}$=1/2 magnets. For Na$_2$IrO$_3$, the total magnetic moment (1 $\mu_B$) of the $J_{\text{eff}}$=1/2 state is composed of the dominant orbital moment (2/3 $\mu_B$) and the spin moment (1/3 $\mu_B$) [26]. Our previous study showed that the directions of the spin and orbital moments in the $J_{\text{eff}}$=1/2 magnets, including Na$_2$IrO$_3$, $\alpha$-Li$_2$IrO$_3$ and $\alpha$-RuCl$_3$, will seriously deviate from each other in the first-principles calculations if the directions of the orbital moments are not constrained [27]. To extract the magnetic interaction parameters of the $J_{\text{eff}}$=1/2 magnet, it is necessary to obtain the total energy of magnetic states with the given directions of spin and orbital moments. So it is crucial to constrain the direction of orbital moments in the $J_{\text{eff}}$=1/2 magnets. Note that the widely used energy-mapping method, which only accounts for the spin, is obviously not applicable here.

In this work, we study the magnetic interactions of the $J_{\text{eff}}$=1/2 magnet Na$_2$IrO$_3$ by combining maximally localized Wannier functions (MLWFs) with our newly developed method [27] which constrains the directions of orbital moments. Apart from the previously well-studied dominant NN magnetic interactions, we find that the long-range third NN Ir-Ir bonds have sizable AFM Heisenberg interactions whereas the second NN Ir-Ir bonds have negligible magnetic interactions. By projecting onto the $J_{\text{eff}}$=1/2 Wannier orbitals, we show that the third NN hopping is much stronger than the second NN hopping, consistent with our calculated magnetic interaction parameters, and that the extended nature of the $J_{\text{eff}}$=1/2 state in the honeycomb lattice gives rise to the sizable third NN Heisenberg interaction $J_3$. Based on our calculated results, we propose a minimal $J_1$-$K_1$-$\Gamma_1$-$J_3$ model for Na$_2$IrO$_3$ which well explains the experimental observations: (I) Its magnetic excitations have an intensity peak at 5.6 meV, consistent with the inelastic neutron scattering experiment [11]; (II) The third NN AFM Heisenberg interaction $J_3$ stabilizes the zigzag AFM order; (III) The NN symmetric off-diagonal exchange $\Gamma_1$ accounts for the experimentally observed magnetic easy axis. The present work not only shows our newly developed method is a powerful tool to study the magnetism of materials with non-negligible orbital moments, but also takes a significant step toward understanding the novel magnetism of honeycomb $J_{\text{eff}}$=1/2 magnets.

## II. COMPUTATIONAL METHODS

Our first-principles calculations based on density functional theory (DFT) are performed within the generalized gradient approximation (GGA) according to the Perdew-Burke-Ernzerhof (PBE) parameterization as implanted in the Vienna *Ab initio* Simulation Package (VASP) [28]. We use the projector-augmented wave (PAW) method [29] and an energy cutoff of 500 eV. To describe the electron correlation associated with the Ir 5$d$ electron, we use the rotationally invariant DFT+U method introduced by Liechtenstein [30]. The on-site Coulomb energy U=3.0 eV and the Hund coupling $J_h$=0.6 eV [15] are adopted in the present work. Because the Ir atom has a strong spin-orbit coupling (SOC) $\xi_{SO}$~0.4 eV [26], SOC is included in our all calculations. We use the experimental monoclinic crystal structure with the space group *C*2/*m* [11].

Since Na$_2$IrO$_3$ is a $J_{eff}$=1/2 magnet [14, 16, 19, 22], we adopt our recently developed methodology that constrains the directions of orbital moments [27] so as to take into account the important effects of orbital moments properly. To gain a deep insight into the magnetic interactions, hopping parameters are extracted from the real-space Hamiltonian matrix elements in the $J_{eff}$=1/2 Wannier orbital basis [31, 32], which are obtained by employing the vasp2wannier90 interface combined with the wannier90 tool [33]. To keep the symmetry of Wannier functions, we utilize the one-shot Wannier construction, in which the minimization of Wannier spread is not performed. The magnetic transition temperature of Na$_2$IrO$_3$ is estimated by performing efficient parallel tempering Monte Carlo (MC) simulations [34-36]. A 40×20×1 supercell of the unit cell, which contains 3200 magnetic ions, are used in these MC simulations.

## III. RESULTS

In this section, we first exhibit that the optimal relative angle between orbital and spin moments is fairly small and then confirm by means of our newly developed method that the experimentally observed zigzag AFM structure is truly the magnetic ground state of Na$_2$IrO$_3$. Next we disclose the sizable long-range third NN AFM Heisenberg interaction has its roots in the spatially extended nature of the $J_{eff}$=1/2 state of the honeycomb lattice. Lastly, based on our DFT calculated magnetic interaction parameters, we propose a minimal $J_1$-$K_1$-$\Gamma_1$-$J_3$ model which explains experimental observations of Na$_2$IrO$_3$ well.

### A. Theoretical reproduction of the zigzag AFM ground state of Na$_2$IrO$_3$

Na$_2$IrO$_3$ is a layer honeycomb antiferromagnet. It crystallizes in the monoclinic space group *C2/m* [11] and consists of alternate stacking of Na$_{1/3}$Ir$_{2/3}$O$_2$ and Na layers (FIG. 1a). Na$_{1/3}$Ir$_{2/3}$O$_2$ layers are composed of edge-sharing IrO$_6$ octahedrons and Ir atoms form the honeycomb lattice with Na atoms sitting at the center of the Ir$_6$ hexagon. It has been experimentally shown that Na$_2$IrO$_3$ has a zigzag AFM order (FIG. 1b) below 18 K [10] and that its magnetic easy axis is approximately along the [110] direction in the (*x*, *y*, *z*) coordinates whose cubic *x*-, *y*- and *z*-axes point along the three NN Ir–O bonds in an octahedron (see FIG. 1a) [14].

We first evidence that the directions of the orbital and spin moments of Na$_2$IrO$_3$ have quite slight deviation, strongly supporting the commonly accepted fact that the magnetism of Na$_2$IrO$_3$ is well described by the $J_{eff}$=1/2 state [14, 16, 19, 22]. Due to the trigonal distortion, the $J_{eff}$ = 1/2 state of Na$_2$IrO$_3$ is not pure and mixed with the $J_{eff}$ = 3/2 state. In this case, orbital and spin moments are not necessarily in exactly the same direction and derivate from each other. Hence, the optimal relative angle between orbital and spin moments itself is of fundamental importance. To figure out this optimal relative angle, we fix the orbital moments along four representative and important axes, namely, *x*-, *y*-, *z*- and [110] axes, and rotate the spin moments slightly away from the fixed orbital moments (see more details in Appendix A). As shown in Fig. 1c, 1d, 1f and 1e, the optimal relative angles giving rise to the minima of the energy ΔE caused by the derivation between orbital and the spin moments are pretty small and in order of several degrees. More explicitly, they are 9, 9, 6, and 5 degrees, respectively, when orbital moments are along the *x*-, *y*-, *z*- and [110] axes. Therefore it is reasonable to deem that orbital and spin moments point along the same direction in Na$_2$IrO$_3$, as required by the $J_{eff}$=1/2 state [26]. This is consistent with a recent theoretical study that asserts that Na$_2$IrO$_3$ is located in the relativistic $J_{eff}$=1/2 Mott insulating region [37]. Therefore the directions of orbital and spin moments are constrained exactly the same in our calculations hereafter.

To examine the magnetic ground state of Na$_2$IrO$_3$ theoretically, we consider eight different important magnetic orders. Four representative magnetic orders (FIG. 2a), namely, the FM, Neel AFM, stripe AFM and zigzag AFM orders, have been widely considered in previous studies [9, 24, 27]. Additionally, four more magnetic orders are also taken into consideration (FIG. 2a). The first one is armchair AF, in which the FM chain is propagating along the armchair edge in the honeycomb lattice. The other three magnetic orders are the zigzag-2 AFM, stripe-2 AFM and

armchair-2 AFM. These three magnetic orders are symmetrically non-equal to the above-mentioned zigzag AFM, stripe AFM and armchair AFM because $Na_2IrO_3$ has the $C_{2h}$ point space group rather than the $C_{3v}$ point space group. Note that the zigzag-2 AFM structure is the same as the reported zigzag (X) order in a recent study of $Na_2IrO_3$ [38]. To determine the magnetic ground state, we considered nine different directions along which the magnetic moments align, namely, *a*, *b*, *c*, [100], [010], [001], [110], [101] and [011], for each magnetic order. FIG. 2b shows the energy differences of the above-mentioned eight magnetic orders with respect to the zigzag-[110] order. As can be seen, the zigzag-[110] order has the lowest energy among them as expected. Interestingly, the zigzag-2-[101] order has a comparable total energy with the ground state zigzag-[110] order. To precisely determine the magnetic easy axis of the zigzag AFM order, we performed detailed investigations on its anisotropic energy. Here, the direction of magnetic moments is distinguished by the polar angle $\theta$ and azimuthal angle $\phi$ in the (*x*, *y*, *z*) coordinate system as shown in FIG. 1a. By scanning $\theta$ and $\phi$, we find that $\theta=80°$ and $\phi=225°$ has the lowest energy (FIG. 2c), that is to say, magnetic moments are parallel or antiparallel to the [441] direction. Actually, this direction slightly deviates from the *xy*-plane and approximately points along the [110] direction. The obtained magnetic easy axis therefore is consistent with the experimentally observed one [14]. Note that the conventional DFT calculations cannot find the zigzag AFM in another $J_{eff}=1/2$ magnet $\alpha$-$RuCl_3$ [27] while it found the zigzag AFM with the moments aligned along the local 110 direction in $Na_2IrO_3$ [24], indicating that our method of constraining the direction of orbital moments generally works well for the $J_{eff}=1/2$ magnets.

### B. Magnetic interaction parameters of $Na_2IrO_3$

To explore the nature of the zigzag antiferromagnetism of $Na_2IrO_3$, we consider the general bilinear exchange Hamiltonian, which has been widely adopted in the previous studies of $J_{eff}=1/2$ magnets [17, 19, 20, 23, 24, 39] and has the form of

$$H = \sum_{ij \in \alpha\beta(\gamma)} \left[ J_{ij} \mathbf{S}_i \cdot \mathbf{S}_j + K_{ij} S_i^\gamma S_j^\gamma + \mathbf{D}_{ij} \cdot (\mathbf{S}_i \times \mathbf{S}_j) + \mathbf{S}_i \cdot \mathbf{\Gamma}_{ij} \cdot \mathbf{S}_j \right] \qquad (1).$$

In Eq. (1), *i* and *j* label Ir sites and the pseudospin operator $\mathbf{S}_i$ is a $J_{eff}=1/2$ state localized pseudospin operator with components $S_i^\alpha$ ($\alpha$=*x*, *y*, *z*). Parameters $J_{ij}$, $K_{ij}$ and $\mathbf{D}_{ij}=(D_{ij}^x, D_{ij}^y, D_{ij}^z)$ are the isotropic Heisenberg interaction, bond-dependent Kitaev interaction and Dzyaloshinskii-

Moriya (DM) vector, respectively. The last term is the generalized symmetric off-diagonal exchange [17], which is

$$\mathbf{S}_i \cdot \mathbf{\Gamma}_{ij} \cdot \mathbf{S}_j = \Gamma_{ij}^x \left( S_i^y S_j^z + S_i^z S_j^y \right) + \Gamma_{ij}^y \left( S_i^z S_j^x + S_i^x S_j^z \right) + \Gamma_{ij}^z \left( S_i^x S_j^y + S_i^y S_j^x \right) \quad (1.1).$$

In this model, we consider the magnetic interactions up to the third NN Ir-Ir pairs. Every Ir-Ir bond is labeled by one pseudospin direction γ=(x, y, z) (see FIG. 1b) and other two directions α and β [17]. For convenience, hereafter, the magnitude of the $J_{eff}$=1/2 state localized magnetic moment is absorbed into the magnetic interaction parameters $J_{ij}$, $K_{ij}$, $\mathbf{D}_{ij}$ and $\mathbf{\Gamma}_{ij}$ so that the $J_{eff}$=1/2 state localized pseudospin $\mathbf{S}_i$ and $\mathbf{S}_j$ are unit vectors.

Our calculated magnetic interaction parameters are listed in Table *I*. These magnetic interaction parameters are calculated by means of the variant from our previous four-state method [40] and its details are given in the Appendix B. We see that the NN magnetic interactions are dominant. The NN Kitaev interactions are FM, consistent with the previous *ab initio* study result [22], and dominate over other kinds of magnetic interactions by almost one order in magnitude. Although the NN x-/y-bonds and the NN z-bond have similarly strong FM Kitaev interactions, their AFM Heisenberg interactions are different somewhat. Our calculations indicate that the symmetric off-diagonal exchanges of the NN x-, y- and z-bonds have three sizable components. One of them is AFM while the other two are FM. This is different from the originally proposed one-component symmetric off-diagonal exchange in the Ref. [17]. Note that it is assumed in the Ref. [17] that the $Ir_2O_{10}$ cluster of $Na_2IrO_3$ has the $D_{2h}$ point symmetry, so that the symmetric off-diagonal exchange has only one component. The $IrO_6$ octahedrons in that cluster actually have tilts and rotations, however. Therefore the $Ir_2O_{10}$ cluster has the $C_{2h}$ point symmetry and DFT calculated symmetric off-diagonal exchanges have three components, which is indeed in accord with the case discussed in the supplementary material of Ref. [17]. In addition to the above findings, we also find that the third NN Heisenberg interactions are unexpectedly sizable while the second NN magnetic interactions are extremely weak compared with the NN ones. It is surprising that the third NN AFM Heisenberg interactions are even comparable to the NN ones since the bond distances of the third NN Ir-Ir pairs are about twice as long as those of the NN Ir-Ir pairs (see Table *I*). The underlying physical reasons for such results will be discussed later. Lastly, the NN Ir-Ir pairs have exactly vanishing DM interactions as they have inversion symmetry. For second NN and the third

NN Ir-Ir pairs, their DM interactions are all extraordinarily weak. Considering those, DM interactions are not included further in the following discussions.

Using our calculated magnetic interaction parameters, we perform efficient exchange Monte Carlo (MC) [34-36] simulation and well reproduce that the magnetic ground state is the zigzag AFM order and its magnetic easy axis is almost along the [110] direction, which is consistent with experimental observations [14]. Moreover, our MC simulation shows that the magnetic transition temperature is about 18.9 K, quite closed to the experimentally measured $T_N$ = 15.3 K [11]. This result thus further rationalizes our calculated magnetic interaction parameters using our newly proposed methods.

### C. $J_{eff}$=1/2 Wannier orbitals of $Na_2IrO_3$

To reveal why the long-range third NN Heisenberg interactions are sizable but the second NN magnetic interactions are so weak in $Na_2IrO_3$, we construct four $J_{eff}$=1/2 Wannier orbitals in the primitive cell of $Na_2IrO_3$ [41]. In the cubic crystal field, $J_{eff}$=1/2 states are in the form of [25]

$$|J_{eff}=1/2,1/2\rangle = \frac{1}{\sqrt{3}}|xy,\uparrow\rangle + \frac{i}{\sqrt{3}}|zx,\downarrow\rangle + \frac{1}{\sqrt{3}}|yz,\downarrow\rangle \quad (2.1),$$

and

$$|J_{eff}=1/2,-1/2\rangle = \frac{1}{\sqrt{3}}|xy,\downarrow\rangle + \frac{i}{\sqrt{3}}|zx,\uparrow\rangle - \frac{1}{\sqrt{3}}|yz,\uparrow\rangle \quad (2.2).$$

According to Eq. (2.1), the $|J_{eff}$=1/2, 1/2> Wannier orbital consists of three components, namely, a real part of spin-up, an imaginary part of spin-down and a real part of spin-down. Similarly, the $|J_{eff}$=1/2, -1/2> Wannier orbital (see Eq. (2.2)) is composed of three components as well, namely, a real part of spin-down, an imaginary part of spin-up and a real part of spin-up. To construct the desired Wannier orbitals using MLWFs, we chose an energy window from -0.3 eV to 0.2 eV in which four isolated bands are included, as shown in the DFT+SOC calculated band structure (FIG. 3a). The red dashed-dotted lines are the Wannier-interpolated four bands, which well reproduce the DFT calculated bands, indicating that $J_{eff}$=1/2 Wannier orbitals are well constructed.

The $J_{eff}$=1/2 Wannier orbitals are spatially widely extended, being distributed over the three NN Ir atoms. FIGS. 3c to 3e show the spatial distribution of the three components of the $|J_{eff}$=1/2, 1/2> Wannier orbital for the reference Ir-0 atom. As expected, the Wannier centers of the three components of this Wannier orbital are located near the reference Ir-0 site. In addition, appreciable tails show up distributing over the three NN Ir sites of the reference Ir-0 atom (labeled by Ir-1, Ir-2 and Ir-3 in FIG. 3c). Consequently, for a specified third NN Ir-Ir pair, their two $|J_{eff}=1/2, 1/2\rangle$ Wannier orbitals are closely connected by their tails (see FIG. 3b). Due to distortion of the IrO$_6$ octahedrons, the $|J_{eff}=1/2, 1/2\rangle$ Wannier orbital is slightly contaminated by the extra imaginary part of the spin-up (FIG. 3f). Likewise, The $|J_{eff}=1/2, -1/2\rangle$ Wannier orbital, is widely distributed over the three NN Ir atoms, but is slightly contaminated by the extra imaginary part of the spin-down. Note that the spatially extended nature of the $J_{eff}=1/2$ Wannier orbitals is also reported in the previous study of the honeycomb $\alpha$-Li$_2$IrO$_3$ [42].

In the $J_{eff}=1/2$ Wannier orbital manifold, the model Hamiltonian is [42, 43]

$$H_{eff} = \sum_{\langle i,j \rangle \sigma} t_{n1}^{r_{ij}} c_{i,\sigma}^{\dagger} c_{j,\sigma} + \sum_{\langle\langle i,j \rangle\rangle \sigma\sigma'} t_{n2;\sigma\sigma'}^{r_{ij}} c_{i,\sigma}^{\dagger} c_{j,\sigma'} + \sum_{\langle\langle\langle i,j \rangle\rangle\rangle \sigma\sigma'} t_{n3}^{r_{ij}} c_{i,\sigma}^{\dagger} c_{j,\sigma} \qquad (3).$$

In Eq. (3), hopping parameters are represented in terms of Pauli matrices [43], such as

$$t_{\sigma\sigma'}^{r_{ij}} = C_{r_{ij}}^{0} \delta_{\sigma\sigma'} + \mathbf{C}_{r_{ij}} \cdot \mathbf{\tau}_{\sigma\sigma'} \qquad (4),$$

where $\tau=(\sigma_x, \sigma_y, \sigma_z)$ is the vector of the Pauli matrices, $\mathbf{C}=(C_x, C_y, C_z)$, $\sigma, \sigma' = \pm$ represent the $J_{eff}=1/2$ Wannier orbitals, and the parameter $r_{ij}$ is the displacement vector between two different Ir sites $i$ and $j$. Considering the direct exchange process between two different Wannier orbitals of the $J_{eff}=1/2$ state, the magnetic interaction parameters can be estimated directly from [43]

$$J_{ij} = \frac{4}{U}\left(C_{ij}^0 C_{ji}^0 - \mathbf{C}_{ij} \cdot \mathbf{C}_{ji}\right) \qquad (5.1),$$

$$\mathbf{D}_{ij} = -\frac{4I}{U}\left(C_{ji}^0 \mathbf{C}_{ij} - C_{ij}^0 \mathbf{C}_{ji}\right) \qquad (5.2),$$

and

$$\Gamma_{ij} = \frac{4}{U}\left(\mathbf{C}_{ij}\mathbf{C}_{ji} + \mathbf{C}_{ji}\mathbf{C}_{ij}\right) \qquad (5.3).$$

In Eq. (5.2), the italic letter $I$ is the imaginary unit.

Surprisingly, the third NN hopping is stronger than that of the NN and second NN. It is shown in Table *II* that the third NN hopping parameters are about three times as large as those of the second NN. Because the direct exchange process dominantly contributes to the magnetic interactions of the second NN and the third NN Ir-Ir pairs, we can estimate, based on Eq. (5.1) that the Heisenberg interaction of the third NN Ir-Ir pairs should be stronger by approximate one order of magnitude compared to those of the second NN Ir-Ir pairs. Such estimation agrees with the DFT calculated NN and third NN Heisenberg interaction parameters (see Table *I*). Besides, the second NN Ir-Ir pairs have similarly small parameters $C^0_{r_{ij}}$ and $C_{r_{ij}}$, so their symmetric off-diagonal exchanges and DM interactions are weak, which is also consistent with our DFT calculated results. Note that the NN Ir-Ir pairs have strong magnetic interactions although they have small hopping parameters in the $J_{\text{eff}}=1/2$ Wannier orbital manifold. The smaller NN hopping than the NNNN hopping is due to the cancellation between different hopping contributions, including *d-d* and *d-p-d* hopping processes [42]. Since the NN Heisenberg interactions have rather different dependences on their various hopping processes [21, 43], it is not applicable to estimate them based on the hopping parameters between the $J_{\text{eff}}=1/2$ Wannier orbitals.

The spatially extended nature of the $J_{\text{eff}}=1/2$ state in the honeycomb lattice can explain the sizable long-range third NN AFM Heisenberg interaction. We have shown that the third NN Ir-Ir pair has two closely connected $J_{\text{eff}}=1/2$ Wannier orbitals due to their appreciable tails distributing over the three NN sites. So it is acceptable that the third NN Ir-Ir pair has a sizable magnetic interaction. If the tails of these two $J_{\text{eff}}=1/2$ Wannier orbitals were removed, they would be far away from each other. In that case, the third NN Ir-Ir pair should have a weak magnetic interaction. To verify this, we artificially removed the tails of the two $J_{\text{eff}}=1/2$ Wannier orbitals of the third NN Ir-Ir pair (see FIG. 3b) by replacing the four bridging Ir atoms (the Ir atoms labeled by the black and blue Ir-1 and Ir-2 in FIG. 3b) with isovalent Si and Ti atoms. Our DFT calculations show that the third NN Heisenberg interaction significantly reduces to -0.11 meV in the case of Si and to 0.04 meV in the case of Ti from the original 0.82 meV. Such a significant reduction indicates that the tails of the $J_{\text{eff}}=1/2$ Wannier orbital are indeed critical to give rise to the sizable long-range third NN Heisenberg interaction. Because the spatially extended nature of the $J_{\text{eff}}=1/2$ state is characterized by these tails, we come to the conclusion that it leads to the sizable long-range third NN AFM Heisenberg interaction.

### D. Minimal $J_1$-$K_1$-$\Gamma_1$-$J_3$ model of Na$_2$IrO$_3$

The general bilinear exchange model, Eq. (1), has many magnetic interaction parameters, which masks the underlying physics of the zigzag antiferromagnetism of Na$_2$IrO$_3$. Actually, some of them are negligible and can be reasonably ignored, and some can be reasonably merged, too. Consequently, a simplified and concise model can be achieved by taking into account the dominant magnetic interactions. To do this, we put forward a minimal $J_1$-$K_1$-$\Gamma_1$-$J_3$ model based on our understanding and the calculated magnetic interaction parameters. Our minimal model is in the form of

$$H = \sum_{\langle ij \rangle \in \alpha\beta(\gamma)} \left[ J_1 \mathbf{S}_i \cdot \mathbf{S}_j + K_1 S_i^\gamma S_j^\gamma + \Gamma_1 \left( S_i^\alpha S_j^\beta + S_i^\beta S_j^\alpha - S_i^\alpha S_j^\gamma - S_i^\gamma S_j^\alpha - S_i^\gamma S_j^\beta - S_i^\beta S_j^\gamma \right) \right]$$
$$+ \sum_{\langle\langle\langle ij \rangle\rangle\rangle \in \alpha\beta(\gamma)} J_3 \mathbf{S}_i \cdot \mathbf{S}_j \qquad (6).$$

In this model, the second NN magnetic interactions are not taken into consideration since they are much weaker than the NN and the third NN ones. It is worthwhile noting that the off-diagonal symmetric exchange $\Gamma_1$ in our model has three components, which is significantly different from the previous theoretical models. Although the NN (the third NN) x-/y-bonds and z-bond are symmetrically nonequivalent, they are considered to be symmetrically equivalent for simplicity. As for the third NN x-, y- and z-bond Ir-Ir pairs, only their Heisenberg interactions are involved in this model. The magnitudes of the magnetic interaction parameters $J_1$, $K_1$, $\Gamma_1$ and $J_3$ in this minimal model are obtained by averaging the corresponding DFT calculated ones, which are listed in Table I. Our MC simulations of the minimal $J_1$-$K_1$-$\Gamma_1$-$J_3$ model with those parameters show that its magnetic ground state is the zigzag AFM with the magnetic easy axis along the [110] direction and its magnetic transition temperature is 17.4 K, very close to the experimentally observed $T_N$ = 15.3 K. Therefore this model describes the experimentally observed zigzag antiferromagnetism of Na$_2$IrO$_3$ well.

The magnetic excitations of the minimal $J_1$-$K_1$-$\Gamma_1$-$J_3$ model have an intensity peak at 5.6 meV, highly consistent with the inelastic neutron scattering experiment [11]. Because of the breakdown of magnons in the strongly spin-orbital coupled magnet [44, 45], we studied the magnons of Na$_2$IrO$_3$ by numerically calculating dynamical structure factors based on the exact diagonalization (ED) computations. Here the dynamical structure factor at zero temperature is defined as [46]

$$I^{\alpha\beta}(\mathbf{Q},\omega) = -\frac{1}{\pi} \text{Im}\left[ \langle 0 | O^{\alpha\dagger} \frac{1}{\omega + E_0 + i\varepsilon - H} O^\beta | 0 \rangle \right] \qquad (7).$$

In Eq. (7), $|0\rangle$ is the ground state wave function of $H$ with the energy $E_0$ and the operator

$O^\alpha = N^{-1} \sum_r S^\alpha \exp(-i\boldsymbol{Q}\cdot\boldsymbol{r})$. The ground state wave function $|0\rangle$ and energy $E_0$ are calculated by the Lanczos method [46] and the intensity $I(\boldsymbol{Q},\omega) = \sum_{\alpha=x,y,z} I^{\alpha\alpha}(\boldsymbol{Q},\omega)$ is obtained by a continued fraction expansion [45-47]. Similar to previous studies [44, 45], we take into account two different 24-site periodic clusters compatible with the zigzag AFM order, namely, the UC-3×2 (FIG.4a) and $\sqrt{3} \times \sqrt{3}$ -2×2 (FIG. 4b) clusters. The energy scan of the dynamical structure factor (FIG. 4c) shows that the magnetic excitations of the UC-3×2 cluster has an intensity peak at 5.6 meV, consistent with the experimentally measured spin wave intensity peak at 5.0 meV in the inelastic neutron scattering experiment [11]. Actually, the magnetic excitation of the $\sqrt{3} \times \sqrt{3}$ -2×2 cluster has an intensity peak at 2.6 meV (FIG. 4c), which should be in accordance with the potential spin wave intensity peak near 2.0 meV in the inelastic neutron scattering experiment [11]. Note that the latter intensity peak is not well-defined in the inelastic neutron scattering experiment because of the limitation of the instrumental energy resolution [11]. Therefore our theoretical result calls for further experimental magnon measurement to clearly identify the possible *hidden* intensity peak so as to comprehensively unveil the nature of the zigzag antiferromagnetism in Na$_2$IrO$_3$.

The experimentally observed zigzag AFM structure is cooperatively established by the NN symmetric off-diagonal exchange and the third NN AFM Heisenberg interaction. If only the NN FM Kitaev interaction $K_1$ and the NN AFM Heisenberg interactions $J_1$ are considered, MC simulations show the magnetic ground state is the stripe AFM, same as the previous study result [21]. To unravel why Na$_2$IrO$_3$ has the [110]-oriented zigzag AFM order, we determined the preferred magnetic orders in the $\Gamma_1$-$J_3$ plane by fixing the NN Heisenberg interaction $J_1$ and the Kitaev interaction $K_1$. To this end, we use the classical Luttinger-Tisza (LT) method in which the pseudospins are considered to be classical moments and the constant length pseudospin vectors are replaced by the unconstrained vector fields $\vec{\phi}_r$. In this case, the classical Hamiltonian written in the momentum space of the unit cell (containing four Ir atoms) of Na$_2$IrO$_3$ is

$$H_{\text{LT}}(\boldsymbol{k}) = \sum_k \phi^\dagger_{k\mu} \Lambda_{\mu\nu}(\boldsymbol{k}) \phi_{k\nu} \qquad (8).$$

In Eq. (8), the Hamiltonian $\Lambda(\boldsymbol{k})$ is a 12-by-12 matrix which is dependent on the magnetic interactions parameters $J_1$, $K_1$, $\Gamma_1$ and $J_3$ (see the details in Appendix C). FIG. 4d shows the phase diagram obtained by numerically minimizing the Hamiltonian $\Lambda(\mathbf{k})$ in the first Brillouin zone. It is shown by FIG. 4d that the third NN AFM Heisenberg interaction $J_3$ stabilizes the zigzag AFM order. This is reasonable, because the third NN AFM Heisenberg interaction is magnetically satisfied in the zigzag AFM order (see FIG. 1b). Besides, the NN symmetric off-diagonal exchange can determine the magnetic easy axis of the zigzag AFM order: (I) If it is FM, magnetic moments will lie in the *ab* plane; (II) If it is AFM, the magnetic moments will be along the [110] direction, namely, the experimentally observed one. Hence it is the cooperation between the NN symmetric off-diagonal exchange $\Gamma_1$ and the third NN AFM Heisenberg interaction $J_3$ that establishes the experimentally observed zigzag AFM structure.

Quantum fluctuations have almost no significant effect on the preferred magnetic orders obtained by the LT method, expect for the phase boundary. Because the $J_{eff}=1/2$ state is an analog to the S=1/2 state [6], it would have strong quantum fluctuations. To elucidate the effect of quantum fluctuations, we additionally carried out an ED computation on the 24-site UC-3×2 cluster (FIG. 4a). We calculated the static spin-structure factor

$$S(\mathbf{Q}) = \sum_{ij} \langle \mathbf{S}_i \cdot \mathbf{S}_j \rangle \exp\left[i\mathbf{Q} \cdot (\mathbf{r}_i - \mathbf{r}_j)\right] \qquad (9).$$

The dominant magnetic order is determined according to the wave number $\mathbf{Q}=\mathbf{Q}_{max}$ which has a maximum in the static spin-structure factor $S(\mathbf{Q})$. FIG. 4e shows the phase diagram obtained in the ED study. By comparing the phase diagram obtained by the LT method with that obtained by the ED computation, one can see that all of the classical magnetic orders obtained by the LT method are recovered by the ED computation except that their phase boundary positions are different. Most importantly, the ED computation also shows that the NN symmetric off-diagonal exchange and the third NN Heisenberg interaction cooperatively establish the experimentally observed zigzag AFM structure.

## IV. DISCUSSION AND SUMMARY

It is perhaps general that the honeycomb $J_{eff}=1/2$ magnet has a sizable long-range third NN Heisenberg interaction. Here we underline that the sizable long-range third NN Heisenberg interaction in $Na_2IrO_3$ is robust and independent of the choice of the on-site Coulomb energy U (see Appendix D), similar to our previous results in the honeycomb $J_{eff}=1/2$ magnet $\alpha$-$RuCl_3$ [27].

It has been reported that the honeycomb $\alpha$-Li$_2$IrO$_3$ also has relatively strong third NN hopping parameters [42] and its third NN Heisenberg interaction is even stronger than the first NN Heisenberg interaction [20]. We note that there are Ir-Ir pairs with bond distances of about 6.0 Å, which is close to the bond distance of the third NN Ir-Ir pairs in the honeycomb Na$_2$IrO$_3$, in the three-dimensional hyper-honeycomb $\beta$-Li$_2$IrO$_3$ [48] and the stripe-honeycomb $\gamma$-Li$_2$IrO$_3$ [49]. Therefore these two three-dimensional iridates are new platforms to investigate whether the long-range third NN Heisenberg interaction is sizable, and deserve further study.

In summary, we have fully studied the magnetic interactions of the honeycomb Na$_2$IrO$_3$ via our newly developed method and the maximally localized Wannier functions. We find that the long-range third NN Ir-Ir pairs have sizable AFM Heisenberg interactions. We demonstrate that the sizable long-range third NN AFM Heisenberg interaction results from the extended nature of the $J_{\text{eff}}$=1/2 state in the honeycomb lattice. We propose a minimal $J_1$-$K_1$-$\Gamma_1$-$J_3$ model for Na$_2$IrO$_3$ and further show that its magnetic excitations have an intensity peak at 5.6 meV, highly consistent with the inelastic neutron scattering experiment [11]. Our work shows that our newly developed method is powerful to study the magnetism of materials with non-negligible orbital moments, such as $J_{\text{eff}}$=1/2 magnets, and paves a significant step toward understanding the novel magnetism of honeycomb $J_{\text{eff}}$=1/2 magnets.

## ACKNOWLEDGMENTS

This paper was partially supported by the National Natural Science Foundation of China, and the Research Program of Shanghai Municipality, the Ministry of Education, and Fok Ying Tung Education Foundation.

# Figures

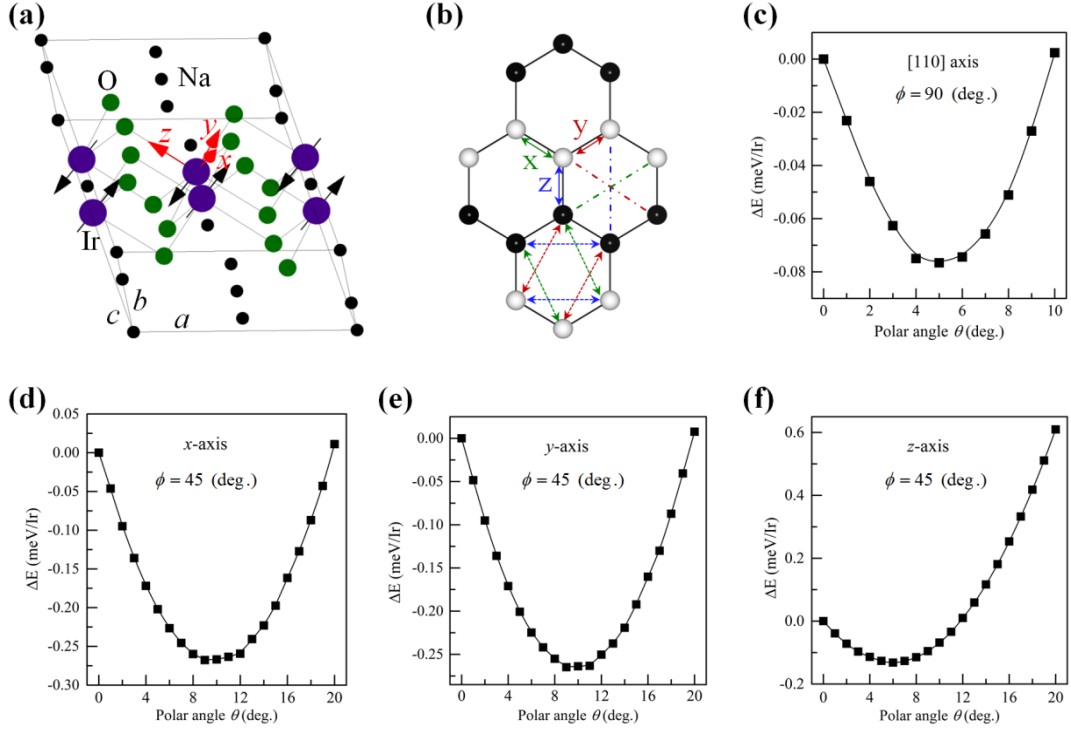

FIG. 1 (Color online) (a) Crystal structure of the honeycomb Na$_2$IrO$_3$. Na, Ir and O atoms are represented by the small black, large purple and dark green medium-size spheres, respectively. The cubic **x**-, **y**- and **z**-axes are the same as the ones used in the Ref. [14] and shown by the red arrows. The zigzag AFM order with magnetic moments along the [110] direction are shown by the black arrows. (b) The NN, second NN and third NN Ir-Ir paths in the Ir atoms sublattice are connected by the solid double-arrowed, dashed double-arrowed, dashed-dotted lines, respectively. The x-, y- and z-bond Ir-Ir paths are distinguished by the green, red and blue color. Black and white spheres have up and down spins, respectively. Dependences of the energy ΔE caused by the derivation between the orbital and the spin moments on the polar angle $\theta$ of spin moment when orbital moment is along the (c) [110], (d) *x*, (e) *y*, and (f) *z* axes.

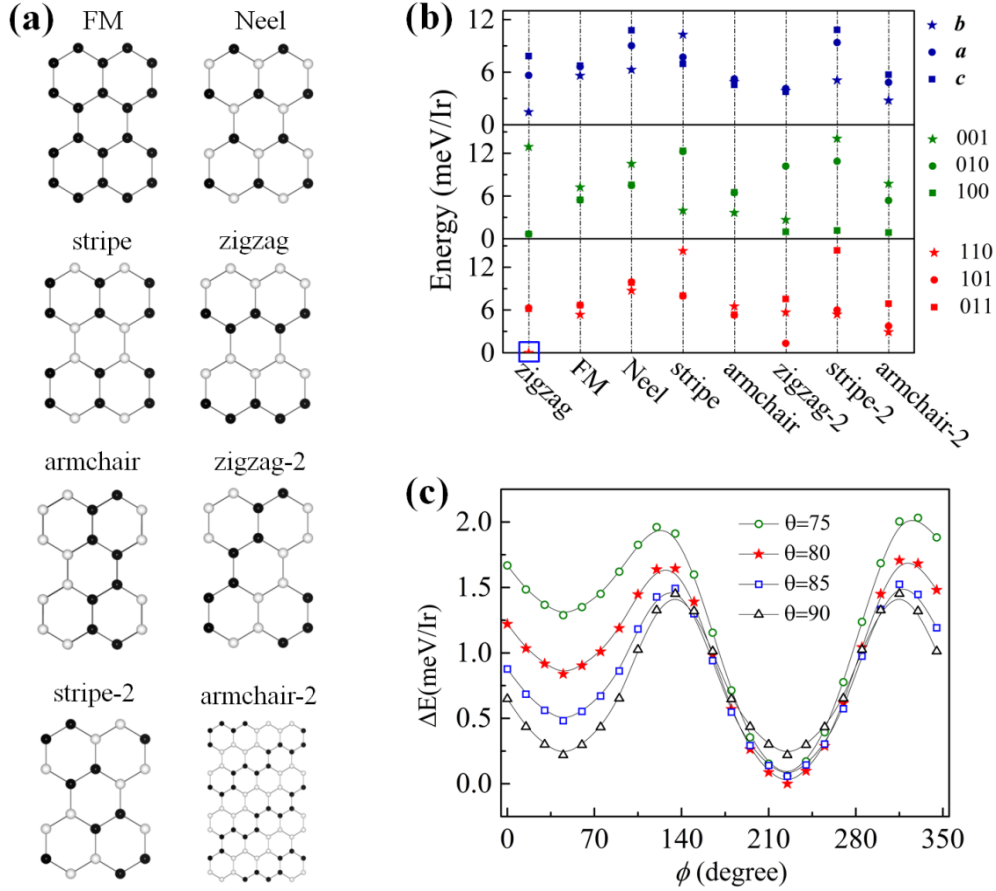

FIG. 2 (Color online) (a) FM, Neel AFM, stripe AFM, zigzag AFM, armchair AFM, zigzag-2 AFM, stripe-2 AFM and armchair-2 AFM magnetic orders. Black and white spheres have up and down spins, respectively. (b) The relative energies of the considered magnetic orders with magnetic moments along various directions. The zigzag AFM order with magnetic moments along the [110] direction (zigzag-[110]) has the lowest energy and is set as the energy reference which is highlighted by the blue rectangle. (c) Energy dependence of the zigzag AFM order on the polar angle $\theta$ and azimuthal angle $\phi$ in the ($x$, $y$, $z$)-coordinate system (see FIG. 1a). The energy of the case ($\theta=80^\circ$, $\phi=225^\circ$) is set as the energy reference. Because the cases of azimuthal $\theta$ ranging from zero to 60$^\circ$ have relatively high energy, they are not shown here.

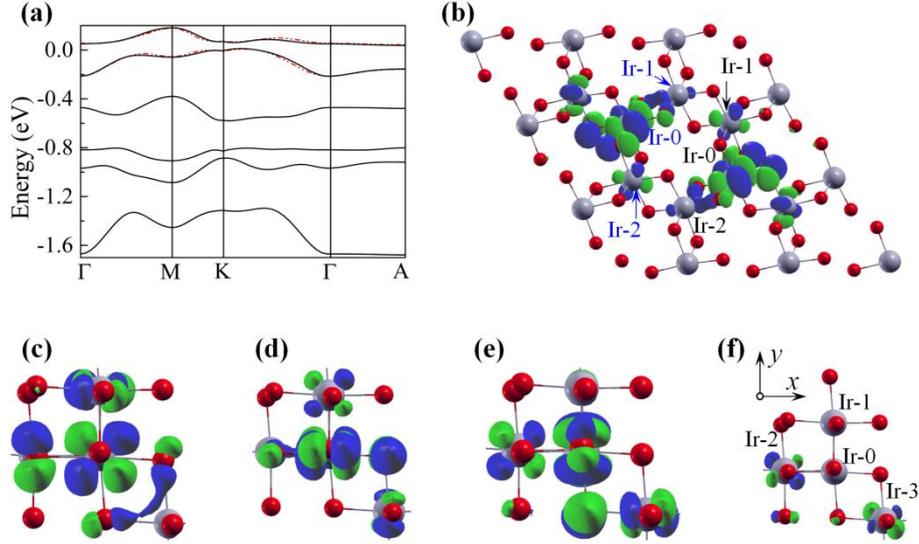

FIG. 3 (Color online) (a) DFT+SOC calculated band structure (black line) of $Na_2IrO_3$ and Wannier-interpolated four bands (the red dashed-dotted lines) near the Fermi level ($E_{Fermi}=0$). (b) Real parts of spin-up of the two $|J_{eff}=1/2, 1/2\rangle$ Wannier orbitals of the third NN Ir-Ir pair (black and blue labels "Ir-0") in the 3×3×1 supercell of $Na_2IrO_3$ primitive cell. (c) Real part of spin-up, (d) imaginary part of spin-down, (e) real part of spin-down and (f) imaginary part of spin-up of the calculated $|J_{eff}=1/2, 1/2\rangle$ Wannier orbital on Ir-0 atom. The three NN Ir atoms of Ir-0 atom are labeled by Ir-1, Ir-2 and Ir-3 as shown in (f). Wannier orbitals in (c)-(f) are viewed along $z$ axis, but they are viewed along the crystallographic $c$ axis in (b). The maximum (minimum) grid values are 7.2 (-7.0) in (c)-(e) and 0.9 (-0.9) in (f). The isosurface value is 1.5 in (b)-(e) and 0.4 in (f).

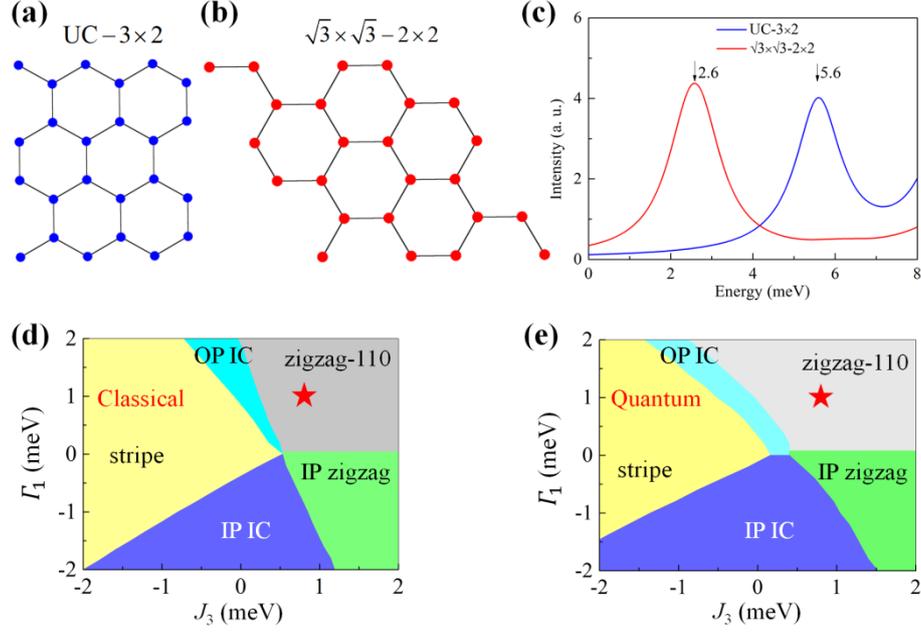

FIG. 4 (Color online) (a) 24-site periodic $UC-3\times2$ and (b) $\sqrt{3}\times\sqrt{3}-2\times2$ clusters. (c) ED calculated energy-dependent magnetic excitation intensity of the minimal $J_1-K_1-\Gamma_1-J_3$ model whose magnetic parameters are listed in Table *I*. A Gaussian broadening of 0.67 meV (FWHM), same as the instrumental energy resolution in the Ref. [11], has been adopted. The intensity peaks of the $UC-3\times2$ (blue) and $\sqrt{3}\times\sqrt{3}-2\times2$ (red) clusters are indicated by the black arrows. Classical (d) and quantum (e) phase diagrams of the minimal $J_1-K_1-\Gamma_1-J_3$ model in the $\Gamma_1-J_3$ plane. The NN Heisenberg $J_1$ and Kitaev interactions $K_1$ are 1.63 and -10.0 meV, respectively. The red stars highlight the specific position ($J_3$ = 0.83 meV, $\Gamma_1$ = 0.90 meV) in the $\Gamma_1-J_3$ plane. The out-of-plane (OP) incommensurate (IC) and the in-plane (IP) IC phase regions can be much richer than the standard coplanar helix states.

Table I. DFT calculated magnetic interaction parameters in units of meV of the general bilinear exchange Hamiltonian, Eq. (1). NNN and NNNN are abbreviated for the second NN and the third NN, respectively. The bond distances of the Ir-Ir pairs are evaluated based on the experimentally measured crystal structure [11]. The last row shows the magnetic interaction parameters present in the minimal $J_1$-$K_1$-$\Gamma_1$-$J_3$ model.

| J Path | | d (Å) | J | K | $\Gamma^x$ | $\Gamma^y$ | $\Gamma^z$ | $D^x$ | $D^y$ | $D^z$ |
|---|---|---|---|---|---|---|---|---|---|---|
| NN | x/y | 3.129 | 1.37 | -9.56 | -1.07 | 1.00 | -1.22 | 0 | 0 | 0 |
|  | z | 3.138 | 2.16 | -11.00 | -0.80 | -0.89 | 0.66 | 0 | 0 | 0 |
| NNN | x/y | 5.425 | 0.08 | -0.14 | 0.01 | 0.06 | 0.07 | -0.12 | -0.13 | 0.00 |
|  | z | 5.427 | 0.16 | -0.16 | 0.06 | 0.06 | 0.15 | 0.15 | 0.05 | 0.17 |
| NNNN | z | 6.257 | 0.82 | 0.14 | -0.01 | 0.01 | 0.04 | 0.00 | 0.01 | -0.03 |
|  | x/y | 6.269 | 0.83 | 0.13 | 0.05 | 0.01 | -0.02 | 0.01 | -0.01 | -0.01 |
| Minimal model | | | $J_1$ 1.63 | | $K_1$ -10.00 | | $\Gamma_1$ 0.90 | | $J_3$ 0.83 | |

Table II. Hopping parameters in units of meV of the NN, second NN (NNN), and third NN (NNNN) Ir-Ir paths in terms of the $J_{\text{eff}}$=1/2 Wannier orbital representation.

| Ir-Ir bond | | $C^0_{r_{ij}}$ | $-iC^x_{r_{ij}}$ | $-iC^y_{r_{ij}}$ | $-iC^z_{r_{ij}}$ |
|---|---|---|---|---|---|
| NN | x | -0.9 | | | |
|  | y | -0.8 | | | |
|  | z | 9.8 | | | |
| NNN | x | -5.8 | 11.4 | 6.3 | -5.3 |
|  | y | -5.8 | -11.1 | 5.8 | 6.31 |
|  | z | -7.3 | 3.7 | -11.2 | 11.6 |
| NNNN | x | -39.0 | | | |
|  | y | -38.8 | | | |
|  | z | -38.3 | | | |

**Appendix A: The optimal relative angle between orbital and spin moments**

To evaluate the optimal relative angle between orbital and spin moments in Na$_2$IrO$_3$, we fix orbital moments along four representative and important axes, namely, $x$-, $y$-, $z$- and [110] axes, and rotate spin moments slightly away from the fixed orbital moments. We define the derivation between orbital and spin moments by a spherical coordinate system in which the polar angle $\theta$ is inclination from the fixed orbital moments and the azimuthal angle $\phi$ lies on the plane perpendicular to the fixed orbital moments. In such definition, the relative angle between orbital and spin moments is the polar angle $\theta$. Note that any angle $\theta$ has extra degree of freedom, namely, the azimuthal angle $\phi$. In order to obtain the correct energy $\Delta E$ caused by the derivation between orbital and the spin moments with a given $\theta$ and $\phi$, we should calculate energies of two different cases. For first case with the energy $E_1(\theta,\phi)$, the orbital moment is fixed at a given axis, for example the $z$ axis, and the spin moment is along the direction with the polar angle $\theta$ and the azimuthal angle $\phi$. For the second case with the energy $E_2(\theta,\phi)$, orbital and spin moments are both in the same direction which is along the vector sum of orbital and spin moments of the first case. Because the general bilinear exchange Hamiltonian Eq. (1) in the main text is anisotropic due to its anisotropic exchange interactions, the second case is actually a reference of the first case. Altogether, the energy $\Delta E$ is calculated by $\Delta E = E_1(\theta,\phi) - E_2(\theta,\phi)$.

The dependences of the energy $\Delta E$ on the polar angle $\theta$ and the azimuthal angle $\phi$ of spin moment (Figure A1) indicate that the optimal relative angle between orbital and spin moments should be found with the azimuthal angle $\phi$ equal to 45 degree when the orbital moment is along the $x$-, $y$- and $z$-axis. However, when orbital moment is along the [110] axis, the optimal relative angle should be found with the azimuthal angle $\phi$ equal to 90 degree. In the main text, we explore in detailed the dependences of the energy $\Delta E$ with the specified azimuthal angles $\phi$ according to the above-mentioned facts. We find that the optimal relative angles between orbital and spin moments are 9, 9, 6, and 5 degrees, respectively, when the orbital moments are along the $x$-, $y$-, $z$- and [110] axes. In principle, orbital moment should scan all the directions to assess the optimal relative angle. Obviously, that is intractable practically. Anyhow, we argue based on the four representative and important directions of orbital moments that the optimal relative angle between orbital and spin moment is rather small and in the order of several degrees.

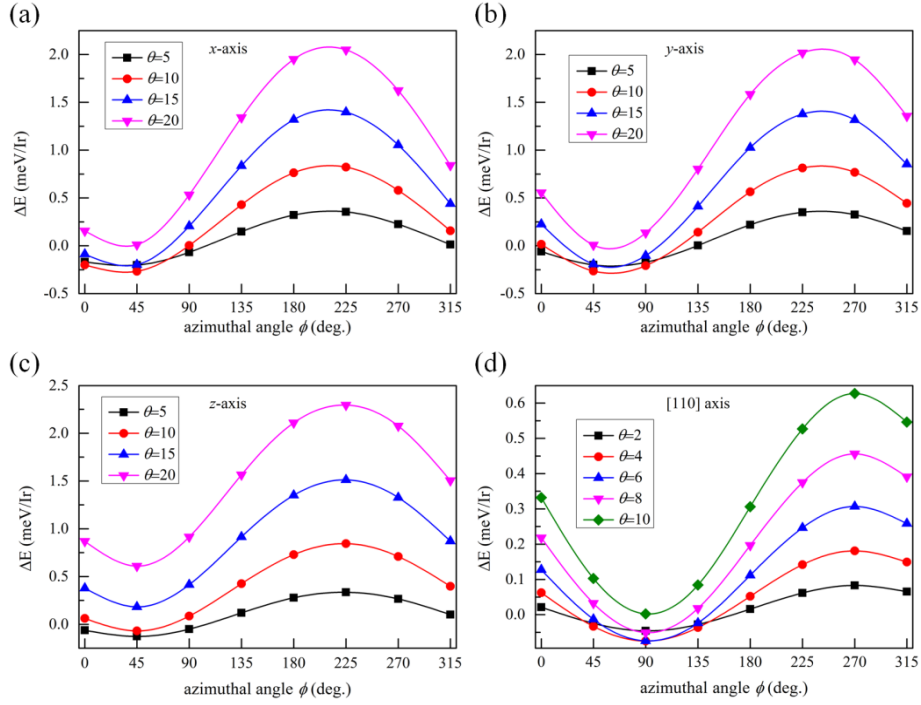

Figure A1. (Color online) Dependences of the energy $\Delta E$ on the polar angle $\theta$ and the azimuthal angle $\phi$ of spin moment when orbital moment is along the (a) $x$, (b) $y$, (c) $z$ and (d) [110] axes.

## Appendix B: Variant of four-state method for calculating magnetic interactions in Hamiltonian Eq. (1)

Here are the details of the variant from our previous four-state method [40] to calculate magnetic interaction parameters in the general bilinear exchange Hamiltonian Eq. (1) in the main text. Another form of the Hamiltonian Eq. (1) is as follows:

$$H = \sum_{ij}\sum_{\alpha\beta} J_{ij}^{\alpha\beta} S_i^{\alpha} S_j^{\beta} \qquad (A.1).$$

In Eq. (A. 1), $i$ and $j$ are the indexes of magnetic sites, and $\alpha$ and $\beta$ run over $x$, $y$ and $z$. Note that the $J_{ij}^{\alpha\beta}$ in the Hamiltonian Eq. (A.1) is a 3-by-3 matrix for a specific $ij$ bond and the exchange parameters in the Hamiltonian Eq. (1) are obtained by appropriate combinations of the elements of the matrix $J_{ij}^{\alpha\beta}$. We take the $ij$ bond belonging to the $xy(z)$ type as an example. In this case, we have

$$J_{ij} = \left(J_{ij}^{xx} + J_{ij}^{yy}\right)/2 \qquad (A.2.1),$$

$$K_{ij} = J_{ij}^{zz} - \left(J_{ij}^{xx} + J_{ij}^{yy}\right)/2 \qquad (A.2.2),$$

$$\Gamma_{ij}^{\alpha} = \left(J_{ij}^{\beta\gamma} + J_{ij}^{\gamma\beta}\right)/2 \qquad (A.2.3),$$

$$D_{ij}^{\alpha} = \left(J_{ij}^{\beta\gamma} - J_{ij}^{\gamma\beta}\right)/2 \qquad (A.2.4).$$

For the $J_{\text{eff}}=1/2$ magnets, $J_{ij}^{xx}$ is equal to $J_{ij}^{yy}$ theoretically in Eq. (A.2.1). Now we show how to calculate any element of the matrix $J_{ij}^{\alpha\beta}$ for a specific $ij$ bond. To this end, we rearrange the items in the Hamiltonian Eq. (A.1) as follows:

$$H = \sum_{\alpha\beta} J_{ij}^{\alpha\beta} S_i^{\alpha} S_j^{\beta} + \sum_{k \neq i,j} \sum_{\alpha\beta} J_{ik}^{\alpha\beta} S_i^{\alpha} S_k^{\beta} + \sum_{k \neq i,j} \sum_{\alpha\beta} J_{kj}^{\alpha\beta} S_k^{\alpha} S_j^{\beta} + \sum_{k \neq i,j; m \neq i,j} \sum_{\alpha\beta} J_{km}^{\alpha\beta} S_k^{\alpha} S_m^{\beta} \qquad (A.3).$$

As a concrete example, we consider the element $J_{ij}^{xy}$. To calculate this element, we set four different magnetic configurations (we denote the magnetic direction of any magnetic site as [$S_x$, $S_y$, $S_z$]) and obtain their corresponding energy to the Hamiltonian Eq. (A.3) as follows:

(1) The magnetic directions of site $i$ and $j$ are [1, 0, 0] and [0, 1, 0], respective, and its energy is

$$E_1 = J_{ij}^{xy} + \sum_{k \neq i,j} \sum_{\alpha} J_{ik}^{x\beta} S_k^{\beta} + \sum_{k \neq i,j} \sum_{\alpha} J_{kj}^{\alpha y} S_k^{\alpha} + \sum_{k \neq i,j; m \neq i,j} \sum_{\alpha\beta} J_{km}^{\alpha\beta} S_k^{\alpha} S_m^{\beta} \qquad (A.3.1);$$

(2) The magnetic directions of site $i$ and $j$ are [1, 0, 0] and [0, -1, 0], respective, and its energy is

$$E_2 = -J_{ij}^{xy} + \sum_{k \neq i,j} \sum_{\beta} J_{ik}^{x\beta} S_k^{\beta} - \sum_{k \neq i,j} \sum_{\alpha} J_{kj}^{\alpha y} S_k^{\alpha} + \sum_{k \neq i,j; m \neq i,j} \sum_{\alpha\beta} J_{km}^{\alpha\beta} S_k^{\alpha} S_m^{\beta} \qquad (A.3.2);$$

(3) The magnetic directions of site $i$ and $j$ are [-1, 0, 0] and [0, 1, 0], respective, and its energy is

$$E_3 = -J_{ij}^{xy} - \sum_{k \neq i,j} \sum_{\beta} J_{ik}^{x\beta} S_k^{\beta} + \sum_{k \neq i,j} \sum_{\alpha} J_{kj}^{\alpha y} S_k^{\alpha} + \sum_{k \neq i,j; m \neq i,j} \sum_{\alpha\beta} J_{km}^{\alpha\beta} S_k^{\alpha} S_m^{\beta} \qquad (A.3.3);$$

(4) The magnetic directions of site $i$ and $j$ are [-1, 0, 0] and [0, -1, 0], respective, and its energy is

$$E_4 = J_{ij}^{xy} - \sum_{k \neq i,j} \sum_{\beta} J_{ik}^{x\beta} S_k^{\beta} - \sum_{k \neq i,j} \sum_{\alpha} J_{kj}^{\alpha y} S_k^{\alpha} + \sum_{k \neq i,j; m \neq i,j} \sum_{\alpha\beta} J_{km}^{\alpha\beta} S_k^{\alpha} S_m^{\beta} \qquad (A.3.4).$$

Based on the Eq. (A.3.1), Eq. (A.3.2), Eq. (A.3.3) and Eq. (A.3.4), we can obtain the element $J_{ij}^{xy}$ in the form of

$$J_{ij}^{xy} = \left(E_1 - E_2 - E_3 + E_4\right)/4 \qquad (A.4).$$

Note that the Eq. (A.4) has the same form as the Eq. (3) in the Ref. [40]. After we obtain the nine elements of the matrix $J_{ij}^{\alpha\beta}$, we can figure out based on Eq. (A.2.1), Eq. (A.2.2), Eq. (A.2.3) and Eq. (A.2.4) the magnetic interaction parameters $J_{ij}$, $K_{ij}$, $\boldsymbol{D}_{ij}$ and $\boldsymbol{\Gamma}_{ij}$ as shown in the Eq. (1) in the main text.

**Appendix C: The classical Luttinger-Tisza Hamiltonian in momentum space**

In the momentum space of the unit cell of $Na_2IrO_3$, which contains four Ir atoms, the classical Hamiltonian $\Lambda(k)$ (see the Eq. (8) in the main text) obtained by the classical Luttinger-Tisa approximation is a 12-by-12 matrix which is of the form

$$\Lambda(k) = \begin{bmatrix} 0 & H_{12} & 0 & H_{14} \\ H_{12}^\dagger & 0 & H_{23} & 0 \\ 0 & H_{23}^\dagger & 0 & H_{34} \\ H_{14}^\dagger & 0 & H_{34}^\dagger & 0 \end{bmatrix} \qquad (A.5).$$

In Eq. (A.1), all elements are 3-by-3 matrices. The nonvanishing matrices are

$$H_{12} = \frac{1}{2}\begin{bmatrix} J_1 & \Gamma_1 & -\Gamma_1 \\ \Gamma_1 & J_1 & -\Gamma_1 \\ -\Gamma_1 & -\Gamma_1 & J_1+K_1 \end{bmatrix} e^{2\pi i k_y} + \frac{1}{2}J_3\left[1+2\cos(2\pi k_x)e^{2\pi i k_y}\right]E_{3\times 3} \qquad (A.6),$$

$$H_{14} = \frac{1}{2}\begin{bmatrix} J_1 & -\Gamma_1 & \Gamma_1 \\ -\Gamma_1 & J_1+K_1 & -\Gamma_1 \\ \Gamma_1 & -\Gamma_1 & J_1 \end{bmatrix} + \frac{1}{2}\begin{bmatrix} J_1+K_1 & -\Gamma_1 & -\Gamma_1 \\ -\Gamma_1 & J_1 & \Gamma_1 \\ -\Gamma_1 & \Gamma_1 & J_1 \end{bmatrix} e^{2\pi i k_x} \qquad (A.7),$$

$$H_{23} = \frac{1}{2}\begin{bmatrix} J_1 & -\Gamma_1 & \Gamma_1 \\ -\Gamma_1 & J_1+K_1 & -\Gamma_1 \\ \Gamma_1 & -\Gamma_1 & J_1 \end{bmatrix} e^{2\pi i k_x} + \frac{1}{2}\begin{bmatrix} J_1+K_1 & -\Gamma_1 & -\Gamma_1 \\ -\Gamma_1 & J_1 & \Gamma_1 \\ -\Gamma_1 & \Gamma_1 & J_1 \end{bmatrix} \qquad (A.8),$$

$$H_{34} = \frac{1}{2}\begin{bmatrix} J_1 & \Gamma_1 & -\Gamma_1 \\ \Gamma_1 & J_1 & -\Gamma_1 \\ -\Gamma_1 & -\Gamma_1 & J_1+K_1 \end{bmatrix} + \frac{1}{2}J_3\left[e^{2\pi i k_y} + 2\cos(2\pi k_x)\right]E_{3\times 3} \qquad (A.9).$$

In Eq. (A.2) and (A.5), $E_{3\times 3}$ is a 3-by-3 unit matrix. Here we set the lattice constant $a$ and $b$ to be unit and ($k_x$, $k_y$) is a point in momentum space. The parameters $J_1$, $K_1$, $\Gamma_1$ and $J_3$ are the magnetic interaction parameters of the minimal $J_1$-$K_1$-$\Gamma_1$-$J_3$ model (see the Eq. (6) in the main text).

**Appendix D: Dependence of exchange parameters of $Na_2Ir_3$ on the on-site Coulomb energy U**

We show the dependence of exchange parameters of $Na_2Ir_3$ on the on-site U. From Table I in the main text and Table A1, one can obtain that, although exchange parameters vary their magnitudes as the Coulomb energy on-site U changes, the relative strengths between the NN,

NNN and NNNN exchange parameters have slight variations. All in all, the third NN Heisenberg interactions are sizable whereas the second NN magnetic interactions are extremely weak compared with the NN ones.

Table A1. DFT calculated magnetic interaction parameters in units of meV of the general bilinear exchange Hamiltonian, Eq. (1) in the case of U=2.5 and U=3.5 eV.

| | | $J$ Path | $d$ (Å) | $J$ | $K$ | $\Gamma^x$ | $\Gamma^y$ | $\Gamma^z$ | $D^x$ | $D^y$ | $D^z$ |
|---|---|---|---|---|---|---|---|---|---|---|---|
| U=2.5 | NN | x/y | 3.129 | 1.60 | -10.73 | -1.14 | 1.14 | -1.29 | 0 | 0 | 0 |
| | | z | 3.138 | 2.48 | -12.42 | -0.83 | -0.97 | 0.70 | 0 | 0 | 0 |
| | NNN | x/y | 5.425 | 0.08 | -0.18 | -0.02 | 0.07 | 0.05 | -0.12 | -0.15 | -0.05 |
| | | z | 5.427 | 0.20 | -0.22 | 0.05 | 0.07 | 0.15 | 0.21 | 0.07 | 0.17 |
| | NNNN | z | 6.257 | 1.02 | 0.15 | 0.01 | 0.01 | 0.07 | -0.02 | 0.02 | 0.00 |
| | | x/y | 6.269 | 1.03 | 0.15 | 0.07 | 0.01 | 0.01 | -0.01 | -0.01 | 0.02 |
| | | $J$ Path | $d$ (Å) | $J$ | $K$ | $\Gamma^x$ | $\Gamma^y$ | $\Gamma^z$ | $D^x$ | $D^y$ | $D^z$ |
| U=3.5 | NN | x/y | 3.129 | 1.13 | -8.50 | -0.97 | 0.90 | -1.10 | 0 | 0 | 0 |
| | | z | 3.138 | 1.86 | -9.78 | -0.71 | -0.83 | 0.69 | 0 | 0 | 0 |
| | NNN | x/y | 5.425 | 0.07 | -0.11 | -0.00 | 0.06 | 0.03 | -0.08 | -0.11 | -0.02 |
| | | z | 5.427 | 0.13 | -0.12 | 0.04 | 0.05 | 0.10 | 0.13 | 0.04 | 0.12 |
| | NNNN | z | 6.257 | 0.66 | 0.13 | 0.01 | 0.01 | 0.07 | -0.01 | 0.01 | 0.00 |
| | | x/y | 6.269 | 0.68 | 0.12 | 0.06 | 0.01 | 0.02 | -0.00 | -0.00 | 0.01 |